\newcommand{\blind}{1}

\documentclass[12pt]{article}
\usepackage{amsmath,amsthm,amssymb}
\usepackage{enumerate}
\usepackage{graphicx}
\usepackage{natbib}
\usepackage{url}
\usepackage{multirow}
\usepackage{xr}

\allowdisplaybreaks

\newcommand{\var}{\mathrm{var}}
\renewcommand{\hat}{\widehat}

\newcommand{\calW}{{\cal W}}

\newcommand{\calD}{{\mathcal D}}
\newcommand{\cbD}{\boldsymbol{\mathcal{D}}}

\newcommand{\calC}{{\mathcal C}}

\def\bse{\begin{eqnarray*}}
\def\ese{\end{eqnarray*}}
\def\be{\begin{eqnarray}}
\def\ee{\end{eqnarray}}
\def\bsq{\begin{equation*}}
\def\esq{\end{equation*}}
\def\bq{\begin{equation}}
\def\eq{\end{equation}}

\def\var{\mathrm{var}}
\def\cov{\mathrm{cov}}
\def\corr{\mathrm{corr}}

\def\CV{\mathrm{CV}}
\def\wh{\widehat}
\def\wt{\widetilde}

\def\n{\nonumber}

\def\BIF{\mbox{BIF}}
\def\BLUP{\mbox{BLUP}}

\def\CONVEN{{\scriptscriptstyle\mathrm{CONV}}}
\def\GEE{\mathrm{GEE}}

\def\QIC{\mathrm{QIC}}
\def\QR{\mathrm{QR}}

\def\MA{\mathrm{MA}}
\def\MACV{\mathrm{MACV}}

\def\SAR{\mathrm{SAR}}
\def\SEAL{\mathrm{SEAL}}
\def\cov{\mathrm{cov}}

\def\corr{\mathrm{corr}}

\def\argmin{\mathrm{argmin}}

\def\train{\scriptscriptstyle \mathrm{train}}
\def\test{\scriptscriptstyle \mathrm{test}}

\def\bb{{\boldsymbol\beta}}

\def\veta{{\boldsymbol\eta}}
\def\beps{{\boldsymbol\varepsilon}}
\def\eps{{\boldsymbol\epsilon}}

\def\0{{\bf 0}}
\def\1{{\bf 1}}

\def\A{{\bf A}}
\def\H{{\bf H}}
\def\h{{\bf h}}

\def\V{{\bf V}}
\def\f{{\bf f}}
\def\g{{\bf g}}

\def\V{{\bf V}}
\def\w{{\bf w}}
\def\X{{\bf X}}
\def\x{{\bf x}}
\def\I{{\bf I}}
\def\J{{\bf J}}

\def\br{{\bf r}}

\def\Y{{\bf Y}}

\def\Z{{\bf Z}}

\def\bUpsilon{{\boldsymbol \Upsilon}}
\def\bdel{{\boldsymbol \delta}}

\def\bmu{{\boldsymbol \mu}}
\def\wh{\widehat}
\def\wt{\widetilde}
\def\var{\mathrm{var}}
\def\cov{\mathrm{cov}}
\def\corr{\mathrm{corr}}
\def\Normal{\hbox{Normal}}

\def\sumi{\mathop{\sum}\nolimits_{i=1}^n}
\def\sums{\mathop{\sum}\nolimits_{s=1}^{{S}}}
\def\sumj{\mathop{\sum}\nolimits_{j=1}^{n_i}}

\def\s{{[s]}}
\def\ii{{[-i]}}

\def\tt{^{\top}}

\def\Pan{{\mbox{\tiny\rm Pan}}}
\def\Imori{{\mbox{\tiny\rm Imori}}}
\newtheorem{theorem}{\underline{\bf Theorem}}
\newtheorem{corollary}{\underline{\bf Corollary}}
\newtheorem{remark}{\underline{\bf Remark}}

\theoremstyle{definition}
\newtheorem{example}{Example}
\newtheorem{Design}{Design}[section]

\newtheorem{Def}{\underline{\bf Definition}}
\newtheorem{assumption}{\underline{\bf Assumption}}
\def\bse{\begin{eqnarray*}}
\def\ese{\end{eqnarray*}}
\def\be{\begin{eqnarray}}
\def\ee{\end{eqnarray}}
\def\bsq{\begin{equation*}}
\def\esq{\end{equation*}}
\def\bq{\begin{equation}}
\def\eq{\end{equation}}
\def\var{\hbox{var}}
\def\cov{\hbox{cov}}
\def\corr{\hbox{corr}}

\def\wh{\widehat}

\def\wt{\widetilde}

\def\n{\nonumber}

\def\bth{{\boldsymbol\theta}}

\def\bmu{\boldsymbol\mu}

\def\A{{\bf A}}

\def\V{{\bf V}}
\def\g{{\bf g}}

\def\f{{\bf f}}
\def\h{{\bf h}}

\def\I{{\bf I}}

\def\U{{\bf U}}
\def\u{{\bf u}}
\def\v{{\bf v}}

\def\w{{\bf w}}
\def\X{{\bf X}}
\def\H{{\bf H}}

\def\x{{\bf x}}

\def\Y{{\bf Y}}

\def\Z{{\bf Z}}

\def\bSig{{\bf \Sigma}}

\def\Normal{\mathrm{Normal}}

\def\Equal{\mathrm{Equal}}

\def\squarebox#1{\hbox to #1{\hfill\vbox to #1{\vfill}}}

\def\bpsi{{\boldsymbol \psi}}

\def\0{{\bf 0}}
\def\2nd{\mathrm{2nd}}

\def\mC{\mathcal{C}}

\def\var{\hbox{var}}
\def\cov{\hbox{cov}}
\def\corr{\hbox{corr}}

\def\Normal{\hbox{Normal}}

\def\wh{\widehat}
\def\wt{\widetilde}

\def\boxit#1{\vbox{\hrule\hbox{\vrule\kern6pt\vbox{\kern6pt#1\kern6pt}\kern6pt\vrule}\hrule}}

\def\s{{(s)}}

\newcommand{\E}{{\mathbb{E}}}
\def\bSig{{\bf \Sigma}}

\def\bSig{{\bf \Sigma}}

\def\ulim{\mathrm{\overline{\lim}}}
\usepackage{titlesec} 

\begin{document}

\def\spacingset#1{\renewcommand{\baselinestretch}%
{#1}\small\normalsize} \spacingset{1}


\if1\blind
{
\title{\bf Unified optimal model averaging  with a general loss function based on cross-validation\footnote{Dalei Yu and Xinyu Zhang are co-first authors. Corresponding author: Hua Liang. This work was supported in part by the National Natural Science
Foundation of China under Grant 12071414, Grant 71925007, Grant 72091212,
Grant 71988101, and Grant 12288201, and in part by the CAS Project for Young
Scientists in Basic Research under Grant YSBR-008.}}
\author{Dalei Yu$^a$, Xinyu Zhang$^b$, and Hua Liang$^{c}$\\
    $^a$Department of Statistics, School of Mathematics and Statistics,\\
     Xi'an Jiaotong University, Xi'an, China\\
    $^b$Academy of Mathematics and Systems Science,\\
    Chinese Academy of Sciences, Beijing, China\\
    $^c$Department of Statistics, George Washington University, Washington, DC}
  \date{}
  \maketitle
} \fi

\if0\blind
{
  \bigskip
  \bigskip
  \bigskip
  \begin{center}
    {\LARGE\bf Unified optimal model averaging  with a general loss function based on cross-validation}
\end{center}
  \medskip
} \fi

\begin{abstract}
Studying unified model averaging estimation for situations with complicated data structures,
we propose a novel model averaging method based on cross-validation ($\MACV$). $\MACV$ unifies a large class of new and existing model averaging estimators and covers a very general class of loss functions. Furthermore, to reduce the computational burden caused by the conventional leave-subject/one-out cross validation,
we propose a SEcond-order-Approximated Leave-one/subject-out ($\SEAL$) cross validation,
which largely improves the computation efficiency. In the context of non-independent and non-identically
distributed random variables, we establish the unified theory for analyzing the asymptotic behaviors of the
proposed $\MACV$ and $\SEAL$ methods, where the number of candidate models is allowed to diverge with sample size.
To demonstrate the breadth of the proposed methodology, we exemplify four optimal model averaging estimators under four important situations, i.e., longitudinal data with discrete responses, within-cluster
correlation structure modeling, conditional prediction in spatial data, and quantile regression with a potential correlation structure.
We conduct extensive simulation studies and analyze real-data examples to illustrate the advantages of
the proposed methods.
\end{abstract}

\noindent%

\if0\blind
\fi
\newpage

\spacingset{1.8}
\section{Introduction}\label{sec:introduction}
\setlength\abovedisplayskip{1pt plus 1pt minus 1pt}
\setlength\belowdisplayskip{1pt plus 1pt minus 1pt}

Having played a prominent role akin to that by model selection, model averaging has been extensively studied in the past two decades \citep{hansen:2007,wan.zhang.zou:2010,lu.su:2014,Zhang:Yu:Zou:Liang:2016,Zhang:Zou:Liang:Carroll:2019}. Model misspecification is unavoidable in practice \citep{Lv:Liu:2014} and in the presence of potential model misspecification, there are growing theoretical
and empirical evidences showing that model averaging can provide more accurate predictions than its
model selection counterparts \citep{lu.su:2014,Zhang:Zou:Liang:Carroll:2019,Peng:Yang:2021}. This makes model averaging a promising option.

The research concerning model averaging in linear regression models for independent observations has grown in the past two decades \citep{hansen:2007,
wan.zhang.zou:2010,liang.zou.ea:2011,hansen.racine:2012, ando.li:2014,Zhang:Zou:Liang:Carroll:2019}. In the situation where the data structure becomes more complicated, \citet{zhang.wan.ea:2013} investigated the behavior of jackknife model averaging in linear regression for dependent data. \citet{Zhang:Yu:Zou:Liang:2016} investigated the optimal model averaging estimation in generalized linear mixed models.
\citet{Liu:Yao:Zhao:2020} studied the optimal model averaging in time series models. Recently, \citet{Feng:Liu:Yao:Zhao:2021} proposed a nonlinear information criterion for model averaging in nonlinear regression models for continuously distributed data.

Various loss functions have been adopted in different areas of data
analysis \citep{li2019lfs,Christoffersen:Jacobs:2004,Bartlett:Jordan:Mcauliffe:2006} and various model averaging
methods have been established based on different loss functions \citep{Zhang:Yu:Zou:Liang:2016,lu.su:2014}. In the
most of existing studies regarding model averaging, under different data settings, by adopting different loss
functions, models, or estimation methods, specific model averaging approaches can be developed in a case-by-case
manner. Given the promising properties and popularity of model averaging,  additional efforts in this direction are still in great demand. However, committing to such case-by-case studies is not only theoretically challenging but also inconvenient.

Ideally, we hope to develop a general model averaging framework that can deal with a broad range of data structures under various loss functions and is applicable to various estimating methods. The resultant frame possesses promising asymptotic properties/optimality and computational convenience. However,
developing such a unified model averaging method for a general class of loss functions is not a straightforward extension of existing results.
For instance, the situation of longitudinal data with discrete responses and non-normal spatial data, the full likelihood of the data is either too complicated to be explicitly tractable or unavailable. As a result, we do not have any immediate estimator for the  risk function. This raises difficulties in developing weight choice criteria and the corresponding asymptotic properties for model averaging estimators.  This article aims to establish the unified optimal model averaging and its theory. We immediately face the following three challenges.
\begin{enumerate}
\item[(a)] \emph{Since we focus on general loss functions and do not make any specific distributional assumptions in the development of unified model averaging, there is no immediate estimator for the risk function when the candidate models are misspecified}.

\item[(b)]\emph{In the typical context of non-normal and correlated data,  responses connect the linear predictors via nonlinear functions { in a complicated manner}. Thus, the risk function cannot be expressed in a usual form, such as a linear-quadratic-type expression of the weight vector, and the theoretical instrument for analyzing the properties under such a general form of risk functions remains {unknown}, especially when there is a divergent number of candidate models.}

\item[(c)]\emph{Cross-validation ($\CV$) is usually adopted to develop flexible model selection or averaging instruments. However, conventional leave-one-out based $\CV$ is typically computationally burdensome and this poses challenges to the development of computationally efficient model averaging.
	}
\end{enumerate}

In this article, we make efforts to develop a general model averaging framework and contribute in three folds.
\begin{enumerate}

\item[(i)] \emph{We establish the asymptotic theory for the proposed model averaging estimator under two scenarios: (1) all the candidate models are misspecified { in the exact sense (the notion of exact sense is given in Definition \ref{def:miss:exact})}, and (2) at least one of them is correctly specified in the exact sense. These results are then extended to the situation with a divergent number of candidate models.%
}
\item[(ii)]  \emph{
We further propose a second-order-approximated leave-one/subject-out ($\SEAL$) method which significantly reduces the computing time and is asymptotically equivalent to the conventional $\CV$. }

\item[(iii)] \emph{ We {exemplify} four novel optimal model averaging estimators in the context of
    longitudinal data with discrete responses, modeling the within-cluster correlation, conditional
    prediction in spatial data, and quantile regression in the presence of a potential correlation structure, which have not been investigated in the model averaging literature}.

\end{enumerate}

The remaining part of the paper is arranged as follows. Section \ref{sec:model} proposes the unified setup of the model averaging based on $\CV$ and within this framework, three new optimal model averaging estimators, i.e., the model averaging estimators for longitudinal data with discrete responses, conditional prediction in spatial data and quantile regression in the presence of a potential correlation structure are investigated accordingly. Section \ref{sec:asy} studies the asymptotic properties of the proposed unified model averaging method when the number of candidate models is assumed to be fixed. These results are then extended to the situation with a divergent number of candidate models in Section \ref{sec:asy:div}. Section \ref{sec:MACV2nd} provides the development and theoretical analysis for the $\SEAL$. Simulation studies are conducted in Section \ref{sec:sim:miss} to assess the finite-sample performance of the proposed method. Section \ref{sec:case:resp} examines a case study to demonstrate the method's usefulness. Section \ref{sec:conclude} concludes. Additional theoretical results, proofs, further discussions, and additional simulations and case studies are also provided in the Supplemental Materials.

\section{Unified model averaging based on $\CV$}\label{sec:model}

Our goal in the current section is to derive the unified formulation of model averaging based on $\CV$.
For the $i$th subject, we have an $n_i$-dimensional response
$\Y_i = (Y_{i1}, \ldots, Y_{in_i})\tt$ and $n_i\times p$ dimensional matrix of covariates $\X_i=(\X_{i1}, \ldots, \X_{in_i})\tt$ for
$i=1,\ldots,n$. Let $\cbD=\{\cbD_1,\ldots,\cbD_n\}$ represents the observable samples with $\cbD_i=\{\Y_i,\X_i\}$.  We assume that $n_i$'s are fixed. When $n_i>1$ for $i=1,\ldots,n$, $\cbD$ is a dataset with multivariate responses. Whereas, when $n_i=1$, $\cbD$ reduces to a common dataset.
 Moreover, denote $\E\{\g(\Y_i)\}=\h(\f_{0,i})$, where $\g:R^{n_i}\to R^{n_i}$ and $\h:R^{n_i}\to R^{n_i}$ are known vector valued-functions, $\f_{0,i}$ is an unknown $n_i\times 1$ vector consisting of parameters of interest and { in the remaining article, we assume that $\|\f_{0,i}\|$'s are bounded above by a positive constant.} This framework is flexible in general. For example, if the mean function of $Y_{ij}$, denoted by $\mu_{ij}$, is of interest, then $\g(\Y_i)=\Y_i$ and $\h(\f_{0,i})=\f_{0,i}=(\mu_{i1},\ldots,\mu_{in_i})\tt$; If the aim is the variance of $Y_{ij}$, say $v_{ij}$, then  $\g(\Y_i) = \{(Y_{i1}-\mu_{i1})^2,\ldots, (Y_{in_i}-\mu_{in_i})^2\}\tt$ and $\h(\f_{0,i})=\f_{0,i}=(v_{i1},\ldots,v_{in_i})\tt$. In addition, as in \cite{lu.su:2014}, consider a one-dimensional case with $\Y_i=Y_i$ and $\f_{0,i}=f_{0,i}$ being the $\alpha$-quantile of $Y_i$, if $Y_i=f_{0,i}+\epsilon_i$, where $\epsilon_i$'s are unobservable error terms. Then, $\g(\Y_i) = 1_{(-\infty,f_{0,i}]}(Y_i)$ and $\h(\f_{0,i}) = P(Y_i\le f_{0,i})= \alpha$, where $1_{\{\cdot\}}(\cdot)$ is the indicator function.

To recover $\f_{0,i}$ from the observable data, we consider $S$ candidate models. We first consider the situation where $S$ is fixed and then extend the results to the situation with divergent $S$. Under the $s$th candidate model ($s=1,\ldots,S$), $\wh \f_{\s,i}$ is the candidate estimator of $\f_{0,i}$ based on $\cbD$. Moreover, let $\Y_i^0$ be the value that needs to be predicted based on $\cbD$ and satisfies that $\E\{\g(\Y_i^0)\}=\E\{\g(\Y_i)\}=\h(\f_{0,i})$. Unlike the standard setup in most existing model selection/averaging literature where $\Y^0 = ({\Y_1^0}\tt,\ldots,{\Y_n^0}\tt)\tt$ is an independent copy of $\Y=(\Y_1\tt,\ldots,\Y_n\tt)\tt$ (e.g. in Section 3.4 of \cite{Konishi:Kitagawa:2007} and \cite{Zhang:Yu:Zou:Liang:2016}),
in the current study, we allow more flexible structure where $\Y^0$ and $\Y$ can be dependent. {This allows us to analyze and improve the performance of prediction in the form of $\E(\Y_i^0\mid \Y)$, see Example \ref{example:SAR} for more details.} Throughout this study, we aim to predict $\Y_i^0$ by constructing a model averaging estimator based on $\wh \f_{(1),i},\ldots,\wh\f_{(S),i}$. Let $\w=(w_1,\ldots,w_S)\tt$ be the vector of weights which belongs to the set $\calW=\{\w\in[0,1]^S:\sums w_s=1\}$. The unified model averaging estimator has a general form \be\wh \f_i(\w)=\sums w_s\wh\f_{\s,i}.\label{eq:MAP}\ee
The model averaging estimator provided in (\ref{eq:MAP}) 
contains a wide class of model averaging estimators and within this framework, we propose three novel model averaging estimators that have never been considered in the literature before.
\begin{example}
\label{example:Gee}
When focusing on discrete longitudinal data, investigators usually only have partial information about the population distribution. To overcome this difficulty, \citet{Liang:Zeger:1986} introduced a generalized estimating equation ($\GEE$) framework, where only the first/second order moment assumptions are imposed. To be specific, assume that $\E(\Y_i) =\bmu_i= (\mu_{i1},\ldots,\mu_{in_i})\tt$. The estimation theory of $\GEE$ has been studied extensively in literature (see for example
\cite{Wang:2011} and \cite{Ratcliffe:Shults:2008}). In practical situations, we impose parametric assumptions on $\mu_{ij}$ and {the correlation structure}, which may be based on different combinations of covariates or working correlation matrices. This leads to multiple candidate models (i.e., candidate $\GEE$s). In the $s$th candidate $\GEE$, $\mu_{ij}$ is assumed to have form $m(\X_{\s,ij}\tt \bb_\s)$, where $m(\cdot)$ is  the pre-specified mean function and $\X_{\s,ij}$  is a $p_s$-dimensional vector which consists of some elements of $\X_{ij}$ and $\bb_\s$ is the corresponding coefficient vector for the $s$th model. Let
$\wh\bb_\s$ be the $\GEE$ estimator of $\bb_\s$ obtained based on $\cbD$. We aim to find the prediction of $\Y^0$, which is an independent copy of $\Y$. This is equivalent to constructing the out-of-sample prediction of $\bmu_i$ and now $\f_{0,i} = \bmu_i$. Under the $s$th $\GEE$, the prediction of $\f_{0,i}$ is $
\wh \f_{\s,i}=  \{m(\X_{\s,i1}\tt\wh\bb_\s),\ldots, m(\X_{\s,in_i}\tt\wh\bb_\s)\}\tt
$ and this yields that $\wh \f_i(\w)= \{\sums w_s m(\X_{\s,i1}\tt\wh\bb_\s),\ldots, \sums w_s m(\X_{\s,in_i}\tt\wh\bb_\s)\}\tt$, where $\wh\bb_\s$ is the estimator obtained by solving the $s$th $\GEE$.
\end{example}

\begin{example}
\label{example:SAR}
Spatial autoregressive ($\SAR$) models \citep{Lee:2004,Martellosio:Hillier:2020} are widely used in geostatistics and economics.  We assume that the data are generated by $\Y = \rho_0\A_0\Y+ \veta_0 + \V$, where $\Y=(Y_1,\ldots,Y_n)\tt$ is the vector of responses, observed in the region under study,  $n$ is the number of spatial units, $\veta_0=(\eta_{0,1},\ldots,\eta_{0,n})\tt$ is an unknown $n\times 1$ vector that contains unit level information, $\A_0 = (a_{0,ij})_{n\times n}$ is the unknown true spatial weight matrix with $a_{0,ii}=0$, $\rho_0$ is the unknown scalar autoregressive parameter and $\V$ is the $n\times 1$ vector of independent disturbance with zero mean and finite variance $\sigma^2$.  Assume that we have $S$ candidate models at hand, where the $s$th candidate model is defined as $\Y = \X_\s\bb_\s +\rho_\s\A_\s\Y + \V_{\s}$, $\X_\s$ is the $n\times p_s$ dimensional matrix of constant regressors, $\bb_\s$ is the corresponding vector of regression coefficients, $\A_\s =(a_{\s,ij})_{n\times n}$ is the working spatial weight matrix with $a_{\s,ii}=0$ (usually specified based on prior knowledge of the investigators) and $\V_{\s}\sim N(\0_{n\times 1},\sigma_\s^2\I_n)$. It is worthwhile noting that the $s$th model is subject to potential misspecification on both the linear predictors and spatial weight matrix, i.e., $\veta_0$ cannot to be written as $\X_\s\bb_\s^0$ for {any} $\bb_\s^0$ and $\A_\s\neq\A_0$. Now, let $\Y^0 = (Y_1^0,\ldots,Y_n^0)\tt$ be another set of spatial units which is unobservable but connects to $\Y$ by $\Y^0 = \veta_0 +\rho_0\A_0\Y + \V^0$, where $\V^0$ is an independent copy of $\V$. This setup allows us to study the predictions for out-of-region units based on observable sample $\Y$. In this scenario, $n_i\equiv1$. The best linear unbiased prediction ($\BLUP$) of $Y_i^0$ is $ \E(Y_i^0\mid\Y)=\eta_{0,i} +\rho_0\sum_{j=1}^n a_{0,ij}Y_j$ and $\f_{0,i}=f_{0,i} =\E\{\E(Y_i^0\mid\Y)\}=\eta_{0,i} +\rho_0\sum_{j=1}^n a_{0,ij}\E(Y_j)$. Therefore, under the $s$th model, the working empirical $\BLUP$ becomes $\wh\f_{\s,i} = \wh f_{\s,i}=\X_{\s,i}\tt\wh\bb_\s +\wh\rho_\s\sum_{j=1}^n a_{\s,ij}Y_j $, where $\X_{\s,i}\tt$ is the $i$th row of $\X_\s$, $\wh\bb_\s$ and $\wh\rho_\s$ are some reasonable estimators (e.g. the maximum likelihood estimators \citep{Lee:2004}, least square estimators \citep{Lee:2002} or adjusted maximum likelihood estimators \citep{Yu:Bai:Ding:2015}) of $\bb_\s$ and $\rho_\s$, respectively. Given $\w$, the working empirical $\BLUP$ of $Y_i^0$ through model averaging is $\wh \f_i(\w) = \wh f_i(\w)=\sums w_s (\X_{\s,i}\tt\wh\bb_\s +\wh\rho_\s\sum_{j=1}^n a_{\s,ij}Y_j)$.
\end{example}

\begin{example}
\label{example:QR}
Providing much more information than the mean regression model \citep{lu.su:2014}, quantile regression ($\QR$) \citep{Koenker:Bassett:1978} plays an important role in statistics and econometrics. Due to the complexity of quantile function, studying model averaging in $\QR$ is difficult in general. In a seminal study, \cite{lu.su:2014} investigated the optimal model averaging based on cross validation and a Mallows' $C_p$-type weight choice criterion, where the observations are assumed to be independent. Recently, \cite{Wang:etal:2023} proposed a frequentist model averaging method for quantile
regression with high-dimensional covariates in the situation where the observations are independent. In the current article, we remove the assumption of independence. We consider that $Y_i^0 = f_{0,i}+\epsilon_i^0$, $i=1,\ldots,n$, where $\eps^0=(\epsilon_1^0,\ldots,\epsilon_n^0)\tt$ is an independent copy of $\eps$ and $\cov(\eps)=\cov(\eps^0)=\bSig$ and $\bSig>0$. We aim to estimate the $\alpha$-quantile of $Y_i^0 $.
We consider $S$ candidate models and the $\alpha$-quantile under the $s$th candidate model is assumed to have the form $f_{\s,i} = \X_{\s,i}\tt\bb_\s$, where $\X_{\s,i}$ is a $p_s$-dimensional sub-vector of covariates vector $\X_i$. Under model $s$, the $\QR$ estimator of $\bb_\s$ is $\wh \bb_\s$ and thus the model averaging estimator of $f_{0,i}$ is $\wh f_i(\w)=\sums w_s \X_{\s,i}\tt\wh\bb_\s$.
\end{example}
To select appropriate weights, we define the loss function as follows:
\be
L_n(\w)=-2 \sum_{i=1}^n\left(\E_{\Y_i^0}\left[Q\left\{\Y_i^0,\wh \f_i(\w)\right\}\right] - \E_{\Y_i}\left\{Q\left(\Y_i,\f_{0,i}\right)\right\}\right),
\label{eq:loss}
\ee
where $-2Q(\cdot,\cdot)$ is any {well-defined} measure of divergence, $\wh \f_i(\w)$ is obtained
based on the observable data $\cbD$,  and the expectation $\E_{\Y_i^0}(\cdot)$ (or $\E_{\Y_i}(\cdot)$) is taken with respect to the probability density function of $\Y_i^0$ (or $\Y_i$). When it is clear from the context, we shall suppress the use of subscripts in the expectations.  This loss function measures the power of $\wh \f_i(\w)$ in predicting $\Y_i^0$. Now we investigate the loss functions in three important practical situations discussed in Examples \ref{example:Gee}--\ref{example:QR}.
\setcounter{example}{0}
\begin{example}[continued]
Under $\GEE$, since there is no explicit form for the likelihood function, the following quasi-likelihood \citep{Severini:Staniswalis:1994,Pan:2001} is employed,
$
Q\left(\br_i, \u_i\right) = a^{-1}(\phi) \sumj \int_{r_{ij}}^{
	u_{ij}
}{ (r_{ij}  - t) v^{-1}(t) dt} $,
where both $\br_i =(r_{i1},\ldots,r_{in_i})\tt$ and $\u_i = (u_{i1},\ldots,u_{in_i})\tt$ are $n_i\times 1$ general vectors. The quasi-likelihood allows us to assess the prediction loss of statistical models in discrete longitudinal data, {without imposing full distributional assumptions}. If $Y_{ij}$ takes value from $\{0,1\}$ with $P(Y_{ij}=1)=\mu_{ij}=1-P(Y_{ij}=0)$, we can set $v(t) = t(1 - t)$ and $a(\phi)=1$. Then,
	\be
	&L_n(\w) &
= -2\sumi\sumj\left[\mu_{ij}\log\left\{\frac{
		\wh f_{ij}(\w) }{1 - \wh f_{ij}(\w) }\right\} + \log\left\{1 - \wh f_{ij}(\w)\right\} \right.\n\\
	&&\quad\left.-\mu_{ij}\log\left(\frac{\mu_{ij}}{1-\mu_{ij}}\right)-\log(1-\mu_{ij}) \right],
	\label{LBernoulli}
	\ee
where $\wh f_{ij}(\w)$ is the $j$th entry of $\wh \f_i(\w)$ defined in Equation (\ref{eq:MAP}).
Moreover, if $Y_{ij}$ takes value from $\{0,1,2,\ldots\}$ and $\E(Y_{ij})=\mu_{ij}>0$, we can set $v(t) = t$ and $a(\phi)=1$. Then, it can be verified that
	\be
L_n(\w)  =  -2\sumi\sumj\left[ \mu_{ij}\log\left\{\wh f_{ij}(\w)\right\} - \wh f_{ij}(\w) -\mu_{ij}\log(\mu_{ij}) + \mu_{ij}\right].    \label{LPoisson}\ee
\end{example}

{\begin{remark}\label{remk:working:idp} In \eqref{LBernoulli} and \eqref{LPoisson}, we employed a ``working independent'' strategy to construct the loss function where the within-cluster correlation structure is ignored. Modeling the correlation or covariance matrix also plays an important role in the literature studying clustered data. In the context of $\GEE$ and longitudinal data analysis, the parameter estimation of $\GEE$ and correlation structure modeling have been studied in a series of seminal papers, e.g., in \cite{Qu:Lindsay:Li:2000}, \cite{Qu:Lindsay:2003} and \cite{Zhou:Qu:2012}, among others. However, with the presence of potential correlation structure, the development of model averaging is not trivial. Inspired by Example \ref{example:Gee}, we provide formal theoretical development of $\MACV$ and numerical studies under our unified framework in Section S.1.3 and Design
S.2 in Section S.4.1 of the Supplementary Materials, respectively.
\end{remark}}

\begin{example}[continued]
Under the $\SAR$ model, we can simply take $
Q\left(r_i, u_i\right) = -(r_i-u_i)^2/2$ and
this leads to
\be
&L_n(\w)&
= \sumi\{f_{0,i} - \wh f_i(\w)\}^2+\sumi\var(Y_i^0)-\sumi \var(Y_i).
\label{LSAR}
\ee
\end{example}

\begin{example}[continued] Under the $\QR$ model, we take $Q(r_i,u_i)= -(r_i-u_i)\{\alpha-1_{(-\infty,0]}(r_i-u_i)\}/2$. Since no specific assumption about the structure of $\bSig$ is imposed, we will work on the ``working-independent'' loss function
\be
L_n(\w)&=&\sumi \E_{Y_i^0}\left(\{Y_i^0-\wh f_i(\w)\}\left[\alpha-1_{(-\infty,0]}\{Y_i^0-\wh f_i(\w)\}\right]\right)\n \\
&&\hspace{2cm} -\sumi\E_{\epsilon_i}\left[\epsilon_i\left\{\alpha-1_{(-\infty,0]}(\epsilon_i)\right\}\right].
\label{LQR}
\ee
\end{example}
It is readily seen from the definition of $L_n(\w)$ in (\ref{eq:loss}) that $\sum_{i=1}^nQ(\Y_i,\f_{0,i})$ only serves as a normalization factor and is unrelated to $\w$. Thus, to minimize $L_n(\w)$ is equivalent to minimizing $-2 \sum_{i=1}^n\E_{\Y_i^0}\left[Q\left\{\Y_i^0,\wh \f_i(\w)\right\}\right]$. This equivalence inspires us to utilize the following $\CV$-based weight choice criterion:
$
\mC_n(\w) =-2 \sumi Q\left\{\Y_i, \wh\f_{i,\ii}(\w)\right\}$,
where $\wh\f_{i,\ii}(\w)=\sums w_s\wh\f_{\s,i}^\ii\n$ and
$\wh\f_{\s,i}^\ii$ is some reasonable prediction of $\f_{0,i}$ based on the data $\X_i\bigcup(\cbD\backslash \cbD_i)$, obtained under the $s$th candidate model. When $n_i=1$ for $1\leq i\leq n$, we refer to the $\CV$ as leave-one-out $\CV$, otherwise, we term it as leave-subject-out $\CV$. The weights are obtained by minimizing $\mC_n(\w)$, i.e.,
$
\wh\w = \argmin_{\w\in\calW}\mC_n(\w)
$
and we term the resulting $\wh\f_i(\wh \w)$ as model averaging estimator by
cross validation or $\MACV$.

\section{Asymptotic properties for fixed $S$}\label{sec:asy}
\subsection{Weak consistency and asymptotic optimality of $\wh\w$}

All limiting processes considered in the current paper correspond to $n\to\infty$, yet $n_i$'s are fixed.
Based on the loss function $L_n(\w)$, we can define the risk function as
$
R_n(\w) = \E_{\Y}\left\{
L_n(\w)\right\}$.
We will establish the asymptotic optimality for $\wh\w$ in the sense of minimizing the prediction risk $R_n(\w)$. To carry this agenda further, we introduce more notations. Use $\left\|\A\right\|$, $\|\A\|_1$ and $\|\A\|_\infty$ to denote the spectral norm, maximum column sum matrix norm and maximum row sum matrix norm \citep{Horn:Johnson:1990} of a general matrix $\A$, respectively. In the remaining part of the paper, denote by $c_0$ and $c_1$ the generic constants that may vary from case to case.

Let $\f_{\s,i}^\ast$ be the limiting value of candidate estimator $\wh\f_{\s,i}$ in the sense that there exists a positive constant $c_1$ such that
$
\sup_{n\ge 1}\max_{1\le i\le n}\E^{1/2}\left\| a_n \left(\wh\f_{\s,i}-\f_{\s,i}^\ast\right)\right\|^2\le c_1$, uniformly for every $s=1,\ldots,S$,
where $ a_n \to\infty$, as $n\to\infty$. In model averaging, $\f_{\s,i}^\ast$ serves as the pseudo-true aggregator \citep{Gospodinov:Maasoumi:2021} which relates closely to the notion of pseudo-true value \citep{White:1982,Lv:Liu:2014}. Moreover, the sequence $ a_n $ can be very different for different scenarios. {In Section S.3.2.1 of the Supplementary Materials, we investigate the behaivor of $a_n$ and demonstrate the relationships between $\wh\f_{\s,i}$, $\f_{\s,i}^\ast$ and pseudo-true values under Examples \ref{example:Gee}, \ref{example:SAR} and \ref{example:QR}.}

Now define $\f_i^\ast(\w) = \sum_{s=1}^S w_s \f_{\s,i}^\ast$,
\be
R_n^\ast(\w)
=-2 \sum_{i=1}^n\E\left[Q\left\{\Y_i,\f_i^\ast(\w)\right\} - Q\left(\Y_i,\f_{0,i}\right)\right],
\label{Rast}
\ee
$\varepsilon_i(\w) = Q\left\{\Y_i,\f_i^\ast(\w)\right\} - Q\left(\Y_i,\f_{0,i}\right)$ and $
\beps(\w) = \{\varepsilon_1(\w),\ldots,\varepsilon_n(\w) \}\tt$.
We state some regularity assumptions for establishing asymptotic properties.
\begin{assumption}
	\label{con:unif:int}
	There exists a positive constant $c_1$ such that for each $s=1,\ldots,S$,
	$\sup_{n\ge1}\max_{1\le i\le n}\E^{{1}/{2}}\| a_n (\wh\f_{\s,i}^\ii - \f_{\s,i}^\ast)\|^{2}\le c_1$,
and $\sup_{n\ge1}\max_{1\le i\le n}\|\f_{\s,i}^\ast\|\le c_1$.
\end{assumption}

\begin{assumption}
		\label{con:w:Mbd} (i) For each $\w\in\calW$, there exists a $u(\w)<\infty$ such that $\left\|\cov\left\{\beps(\w)\right\}\right\|\le u(\w)$, and
		(ii)  there exists a positive constant $c_1$ such that for any $n_{ i}$-dimensional vectors $\f_i$ and $\f_i^\prime$ in the parameter space, $\left| Q(\Z_i,\f_i)-Q(\Z_i,\f_i^\prime)\right|\le K(\Z_i)\left\|\f_i-\f_i^\prime\right\|$, where $\Z_i=\Y_i$ or $\Y_i^0$, and $K(\Z_i)$ is independent of $\f_i$ and $\f_i^\prime$ and satisfies  $\sup_{n\ge 1}\max_{1\le i\le n}\E^{1/2}|K(\Z_i)|^2\le c_1$.
\end{assumption}

Assumption \ref{con:unif:int} concerns the properties of candidate predictions/estimators and essentially means that $\wh \f_{\s,i}$ and $\wh \f_{\s,i}^\ii$ share the same limiting value that is bounded above. In Section \ref{sec:MACV2nd}, we show that if $\wh \f_{\s,i}$ and $\wh \f_{\s,i}^\ii$ can be indexed by a parameter vector, then Assumption \ref{con:unif:int} can be replaced by a group of more straightforward assumptions {that ensures the moment bounds, identifiability and smoothness of the estimating equations}. Further discussions are provided in Section S.3.2.2 of the
Supplementary Materials.  Assumption \ref{con:w:Mbd} regulates the behavior of the loss function. In specific,
Assumption \ref{con:w:Mbd} (i) poses basic pointwise moment condition to $\beps(\w)$, and part (ii) guarantees the stochastic equicontinuity of $Q(\cdot,\cdot)$ with respect to $\f_i$. A similar form of this condition can be found in Condition R3 of \cite{Flynn:Hurvich:Simonoff:2013}, where model selection in misspecified GLM was studied. Unlike in \cite{Flynn:Hurvich:Simonoff:2013}, we avoid assuming that the derivatives of $Q(\cdot,\cdot)$ to exist so that our result applies to a wider class of loss functions (e.g. the check loss considered in Example \ref{example:QR}). Part (ii) of Assumption \ref{con:w:Mbd} also implies the existence of minimizer of $R_n(\w)$ in $\calW$. In particular, for each $\w,\w^\prime\in\calW$,
\be
&\left|\frac{1}{n}R_n(\w)-\frac{1}{n}R_n(\w^\prime)\right|
&\le\frac{2}{n}\sumi \E^{1/2}\left|K(\Y_i^0)\right|^2\max_{1\le s\le S}\E^{1/2}\|\wh \f_{\s,i}\|^2\|\w-\w^\prime\|_1\n\\
&&\le2 c_1^2(1+ a_n ^{-1})\|\w-\w^\prime\|_1,\n%
\ee
which yields that $R_n(\w)$ is a continuous function of $\w$. Therefore, in light of $\calW$ being compact, by the extreme value theorem,  there is a $\w^\ast\in \calW$ such that
$\inf_{\w\in\calW} R_n(\w)=R_n(\w^\ast)$.%

It is seen from (\ref{Rast}) that $R_n^\ast(\w)$ is evaluated at $\f_{\s,i}^\ast$ and in view of the fact that $\wh\f_{\s,i}-\f_{\s,i}^\ast=o_p(1)$, $R_n^\ast(\w)$ essentially serves as the limiting risk. Define
$\mC_n^0(\w) = \mC_n(\w)+ 2\sumi Q\left(\Y_i, \f_{0,i}\right)$, 

where the last term is free from $\w$. It follows that
$\argmin_{\w\in\calW}\calC_n(\w) =\wh\w = \argmin_{\w\in\calW}\mC_n^0(\w)$.
In the remaining part of this paper, we study the asymptotic properties of $\wh\w$ based on $\mC_n^0(\w)$.
In fact, $\calC_n^0(\w)$, $R_n^\ast(\w)$ and $R_n(\w)$ are closely related. To see this, denote $\nu_n =  \sup_{\w\in\calW}\left|\sumi \E\left[Q\{\Y_i,\f_i^\ast(\w)\}-Q\{\Y_i^0,\f_i^\ast(\w)\}\right]\right|$, then we have the following result.

\begin{theorem}\label{thm:unif:convg}
Under Assumptions \ref{con:unif:int} and \ref{con:w:Mbd}, (i) $\sup_{\w\in\calW}\left|\calC_n^0(\w)/n - R_n^\ast(\w)/n\right|=o_p(1)$, as $n\to\infty$; if $\nu_n/n=o(1)$ and $\w^\ast$ is a well-separated minimum of $R_n(\w)$, i.e., for a given $\delta>0$, $\inf_{\{\w\in\calW: \|\w-\w^\ast\|>\delta\}}R_n(\w)>R_n(\w^\ast)$, then $\left\|\wh\w-\w^\ast\right\|=o_p(1)$, as $n\to\infty$.
	\end{theorem}
The proof of the theorem is provided in Section S.2.1 of the Supplemental Materials. The proof is not technically trivial because $\w$ is embedded in $\calC_n^0(\w)$ or $R_n^\ast(\w)$ in a complicated manner and stochastic equicontinuity \citep{Newey:1991} plays a pivotal role in the proof of Theorem \ref{thm:unif:convg}.
In this theorem, we require that $\nu_n/n\to0$, which essentially means that { $\Y^0$ is predictable given $\Y$, i.e.,  $\Y$ and $\Y^0$ should share a certain degree of similarity. This notion is similar to the notion of transformability in transfer learning \citep{Tian:Feng:2023}.} To be specific, when $\Y^0$ is an independent copy of $\Y$ (as in Examples \ref{example:Gee}, \ref{example:QR} and existing literature concerning model selection or model averaging), $\nu_n\equiv0$. Whereas in Example \ref{example:SAR}, under the standard assumptions imposed by \cite{Lee:2002,Lee:2004}, we show that $\nu_n/n=O(h_n^{-1})=o(1)$ in Section S.3.1.3 of the Supplemental Materials.

Theorem \ref{thm:unif:convg} is the first result to establish the weak consistency of $\wh \w$ in unified model averaging and the results obtained in \cite{Zhang:Zou:Liang:Carroll:2019} and \cite{Feng:Liu:Yao:Zhao:2021} can be viewed as special cases of Theorem \ref{thm:unif:convg} under the scenario therein. The assumption that $\w^\ast$ is a ``well-separated minimum'' is weaker than convexity (further discussion of this notion can be found in Section 5.2 of \cite{vanderVaart:2000}). It is also worthwhile noting that if we further assume that $R_n(\w)$ is a strictly convex function of $\w$, then the assumption about ``well-separated minimum'' is satisfied evidently and $R_n(\w)$ attains its unique minimum at $\w^\ast$ in the convex set $\calW$. In this case, by minimizing $\mC_n^0(\w)$, we can recover the \emph{unique} minimizer of $R_n(\w)$, asymptotically.

Theorem \ref{thm:unif:convg} shows that ${n^{-1}}\mC_n^0(\w)$ is a uniformly consistent estimator of the limiting risk ${ n^{-1}}R_n^\ast(\w)$. Therefore, it is expected that when minimizing $\mC_n^0(\w)$, the limiting risk ${ n^{-1}}R_n^\ast(\w)$ is also minimized. Furthermore, since $R_n^\ast(\w)$, as the limiting risk corresponding to $R_n(\w)$, should be close to $R_n(\w)$ for sufficiently large $n$, we expect
that $R_n(\w)$ is also minimized asymptotically. In Theorem \ref{th:opt:MACV}, we will verify this conjecture. To this end, let
$
\wh d_n(\w)=-2{\sum_{i=1}^n\left[Q\left\{\Y_i,\wh\f_{i,\ii}(\w)\right\} - Q\left\{\Y_i,\f_i^\ast (\w)\right\}\right]}/{R_n^\ast(\w)}$
and $\wh q_n(\w) = -2{\sum_{i=1}^n \left[Q\left\{\Y_i,\f_i^\ast(\w)\right\}-Q\left(\Y_i, \f_{0,i}\right)\right]}/{R_n^\ast(\w)}$. We need some additional assumptions. Let $\xi_n=\inf_{\w\in\calW} R_n^\ast(\w)$.
\begin{assumption}
	\label{con:equicon}
	There is a $\wh K_n =O_p(1)$ such that for all $\w^\prime$ and $\w$ in $\calW$,
	$\left|\wh d_n(\w^\prime) - \wh d_n(\w)\right|\le  \wh K_n  \left\|\w^\prime - \w\right\|$
	and
	$\left|\wh q_n(\w^\prime) - \wh q_n(\w)\right|\le  \wh K_n  \left\|\w^\prime - \w\right\|$.
\end{assumption}
\begin{assumption}
		\label{con:miss}
	As $n\to\infty$,	$ \max(n/ a_n ,n^{1/2})/\xi_n=o(1)$ and $\nu_n/\xi_n=o(1)$.
\end{assumption}
Assumption \ref{con:equicon} regulates the behavior of the loss function normalized by $R_n^\ast(\w)^{-1}$ and guarantees the stochastic equicontinuity
of $\wh d_n(\w)$ and $ \wh q_n(\w)$ with respect to $\w$. This is analogous to Assumption 3A in \cite{Newey:1991} and can be replaced by Assumption \ref{con:w:Mbd} (ii) when $\ulim_{n\to\infty}n/\xi_n\to 1/c_0<\infty$ for some positive constant $c_0$ (like in the setup of Theorem 3.3 of \cite{lu.su:2014}). Please see Section S.3.1.2 for more detailed discussions. Assumption \ref{con:miss} regulates the lower bound of limiting risk $R_n^\ast(\w)$ in $\calW$ and requires that $\xi_n$ grows faster than $n/ a_n $, $n^{1/2}$ and $\nu_n$. Similar conditions have also been used in \cite{Zhang:Yu:Zou:Liang:2016} and \cite{ando.li:2014}.
Now we provide some further insights into this condition and we introduce the following definition, which is new in the literature.
\begin{Def}[Definitions for correctly specified models]
\label{def:miss:exact}
\begin{itemize}
\item[(i)] The model $s$ is said to be correctly specified \emph{in the exact sense}, if
$\f_{\s,i}^\ast\equiv\f_{0,i}$ for every $i=1,\ldots,n$, i.e., its pseudo-true aggregator is identical to the true parameter of interest.
\item[(ii)] If the model $s$ is not correctly specified in the exact sense, but satisfies that as $n\to\infty$, $\max_{1\le i\le n}\|\f_{\s,i}^\ast-\f_{0,i}\|\to0$, we say the model is correctly specified \emph{in the asymptotic sense}.
\end{itemize}
\end{Def}

In some of the most commonly used $R_n^\ast(\w)$, we can always find a fixed positive constant $K_1$ such that
\be
&R_n^\ast(\w) &\ge  \frac{1}{K_1}\sumi {\left\|\sums w_s(\f_{\s,i}^\ast - \f_{0,i})\right\|^2 }.
\label{Rnw}
\ee
In S.3.1.4 of the Supplemental Materials, we
verify (\ref{Rnw}) in Examples \ref{example:Gee}, \ref{example:SAR} and \ref{example:QR}. Denote $\bdel_i  = (\f_{(1),i}^\ast - \f_{0,i},\ldots,\f_{(S),i}^\ast - \f_{0,i})\tt$ and
$\bUpsilon_n =(\bdel_1, \ldots, \bdel_n)\tt$.
$\bUpsilon_n$ captures the information that reflects the difference between the pseudo-true aggregator and true data generating process. Denote by $\lambda_{\min}(\A)$ the minimum eigenvalue of a general symmetric matrix $\A$. Note that
$\|\w\|_1 = \sums |w_s|=1$. We have
\be
R_n^\ast(\w) \ge \w\tt\frac{\sumi \bdel_i\bdel_i\tt}{K_1}\w \ge \frac{ \lambda_{\min}(\bUpsilon_n\tt\bUpsilon_n)\|\w\|^2}{K_1}  \ge \frac{\lambda_{\min}(\bUpsilon_n\tt\bUpsilon_n)\|\w\|_1^2}{K_1S} =\frac{\lambda_{\min}(\bUpsilon_n\tt\bUpsilon_n)}{K_1S}\n.
\ee
The above argument indicates that
$
\xi_n= \inf_{\w\in\calW} R_n^\ast(\w) \ge K_1^{-1}S^{-1}\lambda_{\min}(\bUpsilon_n\tt\bUpsilon_n),\n
$
which yields that Assumption \ref{con:miss} holds evidently if $\lambda_{\min}(\bUpsilon_n\tt\bUpsilon_n)$ grows faster than $\max(n/ a_n ,n^{1/2})$ and $\nu_n$.
{ For fixed dimensional parametric regression models, { Assumption \ref{con:miss}} is typically violated if one or more correctly specified models lie within the candidate model set,  in the exact sense. That is, if for $s=s_0$, one has $\f_{{s_0},i}^\ast \equiv \f_{0,i}$, then by the definition of $R_n^\ast(\w)$, we have $\xi_n =\inf_{\w\in\calW} R_n^\ast(\w) =  R_n^\ast(\w_{s_0}^0)=0$, where $\w_s^0$ is the $S\times 1$-dimensional vector whose $s$th element is one and the others are zeros.
However, in the context of nonparametric regression models,  when some fitting models are correctly specified in the asymptotic sense, if the fitting error is dominated by approximation error, then Assumption \ref{con:miss} can still be satisfied (see \cite{Racine:Li:Yu:Zheng:2023} for the definition of correct model in nonparametric setting).
It is also worthwhile noting that Assumption \ref{con:miss} poses a requirement for the divergence rate of $\nu_n$, which} essentially means that given $\Y$, $\Y^0$ is predictable relative to $\lambda_{\min}(\bUpsilon_n\tt\bUpsilon_n)$. In particular, in Examples \ref{example:Gee} and \ref{example:QR}, where $\Y^0$ is an independent copy of $\Y$, $\nu_n/\xi_n\equiv0$. Whereas in Example \ref{example:SAR}, under the standard assumptions imposed by \cite{Lee:2002,Lee:2004},
we show in Section S.3.1.3 of the Supplemental Materials that $\nu_n=O(nh_n^{-1})$, which indicates that Assumption \ref{con:miss} holds when $\lambda_{\min}(\bUpsilon_n\tt\bUpsilon_n)$ grows faster than $n/h_n^{1/2}$. Now we are ready to establish the asymptotic optimality for $\wh \w$.
\begin{theorem}[\bf Asymptotic optimality]
	\label{th:opt:MACV}
Under Assumptions \ref{con:unif:int}--\ref{con:miss}, {as $n\to\infty$,} we have
	\be
	\frac{
		R_n(\wh \w)}{\inf_{\w\in\calW}R_n(\w)}\to 1,
	\label{opt}
	\ee
	in probability.
\end{theorem}
The proof of Theorem \ref{th:opt:MACV} is provided in Section S.2.2 of the Supplemental Materials. Again, the stochastic equicontinuity \citep{Newey:1991} plays a pivotal role in the proof.
Theorem \ref{th:opt:MACV} demonstrates that by minimizing the proposed weight choice criterion $\calC_n(\w)$,
{the obtained} $\wh\w$ is asymptotically optimal in the sense
that its {prediction accuracy measured by the risk function} is asymptotically identical to that based on the best yet infeasible weight vector. This theorem provides the theoretical underpinning for $\MACV$.

\subsection{The asymptotic behavior of $\wh\w$ when there is at least one correctly specified candidate model in the exact sense}
\label{subsec:correctmodel}

In Theorem \ref{th:opt:MACV}, we show that the proposed model averaging method is asymptotically optimal in the sense of minimizing the prediction risk. Playing a pivotal role in the proof of Theorem \ref{th:opt:MACV}, Assumption \ref{con:miss} rules out the situation  { where there is at least one correctly specified candidate model in the exact sense (\cite{Flynn:Hurvich:Simonoff:2013} refers to this as ``true model world'')}. Then, a natural question is that in the ``true model world'', can we consistently identify all the correct models based on $\wh \w$? We answer this question now.

Assume that $D$ is a non-empty set that contains all the labels of
the correctly specified models in the exact sense. Consider a restricted weight set $\overline\calW_D = \{\w=(w_1,\ldots,w_S)\tt\in \calW: w_s\equiv0 {~\text{if}~} s\in D\}$. To study the behavior of $\wh\w$, we assume $D\neq \{1,\ldots,S\}$ and this means that at least one of the candidate models is misspecified. If $D= \{1,\ldots,S\}$, the conclusion in the following
Theorem \ref{th:consist:MACV} holds evidently. We need the following additional assumption.
{\begin{assumption}
		\label{con:dQ2}
		There are two positive constants $c_0$ and $K_1$ such that, (i) for every $\w\in\calW$
		$
		R_n^\ast(\w) \ge \sumi\left\|\f_i^\ast(\w) -  \f_{0,i}\right\|^2/K_1\n$; (ii)
		$
		\inf_{\w\in\overline\calW_D}{\sumi \left\|\sum_{s\notin D}w_s(\f_{\s,i}^\ast - \f_{0,i})\right\|^2}/{n}\ge c_0$ for sufficiently large $n$.
\end{assumption}}

Part (i) of Assumption \ref{con:dQ2} is very mild, and is
satisfied under the standard setup of Examples \ref{example:Gee}, \ref{example:SAR} and \ref{example:QR} (see Section S.3.1.4 of the Supplemental Materials for further discussions).
Part (ii) of Assumption \ref{con:dQ2} means that not all the candidate models are correctly specified in the exact sense and
the corresponding misspecification error does not vanish as $n\to\infty$.

\begin{theorem}[\bf Consistency of $\wh\w$ in identifying the exactly correct models]
	\label{th:consist:MACV}
If Assumptions \ref{con:unif:int}, \ref{con:w:Mbd} and \ref{con:dQ2} are satisfied, then we have $\sum_{s\in D}\wh w_s \to 1$ in probability as
	$n\to\infty$, where $\wh w_s$ is the $s$th entry of $\wh\w$.
\end{theorem}

The proof of Theorem \ref{th:consist:MACV} is provided in Section S.2.3 of the
Supplemental Materials.
This result implies that when there are some
correctly specified models in the exact sense, and the sample size is sufficiently large, our method can successfully identify all these exactly correct models and reduce the weights for the misspecified models to zeros.
Theorems \ref{th:opt:MACV} and \ref{th:consist:MACV} are established under very different situations. The former one is built under the case where there is no correctly specified model, whereas the latter one is derived under the case where at least one of the working models is correctly specified. These two theorems provide theoretical supports for the use of our $\MACV$ under the different practical situations.

\section{Asymptotic properties with divergent $S$}\label{sec:asy:div}
\label{subsec:asy:div}
We now study the asymptotic properties of the proposed method in the context where $S\to\infty$, as $n\to\infty$, which is commonly encountered for high-dimensional regressions or nonparametric regressions. Denote $z_i(\w) = \varepsilon_i(\w)-\E \{\varepsilon_i(\w)\}$. In the remaining part of the current section, we assume that for each $\w\in\calW$, $z_i(\w)$'s satisfy the \emph{mixing condition} given in (S.5). We aim to extend the results in Theorems \ref{thm:unif:convg}, \ref{th:opt:MACV} and \ref{th:consist:MACV} into the scenario with divergent $S$. This type of study is difficult in general. The reason is that for divergent $S$, we cannot adopt equicontinuity directly to analyze the uniform convergence properties of $\calC_n^0(\w)$ over $\calW$ and there is no immediate concentration inequality that can be used in our setting. For this purpose, we establish a variant of Bernstein-type inequality under the situation where observations are non-independent and not necessarily identically distributed by modifying the work of \cite{Modha:Masry:1996} and \cite{Merlevede:etal:2009}, whose version works for stationary time series and stochastic process, respectively.   \citeauthor{Modha:Masry:1996}'s version can be treated as our special case. The variant can be used in the cross-sectional setting with potential spatial or network dependence. The inequality is of interest itself and may be independently useful. We have the following results.
\begin{theorem}\label{thm:unif:convg:div}
Under Assumptions S.1--S.3 of the Supplementary Materials, for each $\w\in\calW$, assume that $z_i(\w)$'s satisfy the mixing condition given in (S.5), then, (i) $\sup_{\w\in\calW} |\calC_n^0(\w)/n  - R_n^\ast(\w)/n |=o_p(1)$, as $n\to\infty$; (ii) if $\nu_n/n=o(1)$ and $\w^\ast$ is a well-separated minimum of $R_n(\w)$, i.e., for a given $\delta>0$, $\inf_{\{\w\in\calW: \|\w-\w^\ast\|_\infty>\delta\}}R_n(\w)>R_n(\w^\ast)$, then $\left\|\wh\w-\w^\ast\right\|_\infty=o_p(1)$, as $n\to\infty$.
	\end{theorem}
The proof of Theorem \ref{thm:unif:convg:div} is provided in Section S.2.4 of the Supplemental Materials. Now we establish the asymptotic optimality of $\wh\w$.
\begin{theorem}[\bf Asymptotic optimality under divergent $S$]
	\label{th:opt:MACV:div}
Under Assumptions S.1, S.2 and S.4 of the Supplementary Materials, for each $\w\in\calW$, assume that $z_i(\w)$'s satisfy the mixing condition given in (S.5), we have, as $n\to\infty$
	$
	{
		R_n(\wh \w)}/{\inf_{\w\in\calW}R_n(\w)}\to 1
	$, in probability.
\end{theorem}
The proof of Theorem \ref{th:opt:MACV:div} is provided in Section S.2.5 of the
Supplemental Materials.
{ Moreover, when there is at least one correctly specified candidate model in the exact sense, we can also develop the consistency theory for $\wh \w$.}
\begin{theorem}[\bf Consistency of $\wh\w$ in identifying the exactly correct models]
	\label{th:consist:MACV:div}
If Assumption \ref{con:dQ2} and S.1--S.3 of the Supplementary Materials are satisfied, for each $\w\in\calW$, assume that $z_i(\w)$'s satisfy the mixing condition given in (S.5), then we have $\sum_{s\in D}\wh w_s \to 1$ in probability as
	$n\to\infty$, where $\wh w_s$ is the $s$th entry of $\wh\w$.
\end{theorem}
Under the same framework as that which proves Theorem \ref{th:consist:MACV}, the proof of Theorem \ref{th:consist:MACV:div} is straightforward using the results established in Theorem \ref{thm:unif:convg:div} and is omitted.

\section{A fast $\MACV$ based on the second-order-approximation}
\label{sec:MACV2nd}
\subsection{A fast algorithm for leave-one/subject-out estimators}
\label{subsec:Est2nd}

To obtain the weight choice criterion $\calC_n(\w)$, one needs to remove $\calD_i$ from the full dataset and conduct parameter estimation or learner training based on $\calD\setminus\calD_i$ for all the $s=1,\ldots,S$ and $i=1,\ldots,n$. The procedure is repeated for $n\times S$ times, which is deemed computationally burdensome, and this restricts the use of $\MACV$ when the candidate models are complicated. The problem becomes much severe when $n$ and/or $S$ are/is large. This severity inspires us to consider a more efficient method to alleviate the computational burden. Developing the fast $\CV$ algorithm is an important issue in machine learning and statistical inference. Focusing on the kernel-based regression, \cite{Debruyne:etal:FastCV:2008} considered a fast algorithm based on evaluating the influence functions at a specific sample distribution to obtain an approximation of the $\CV$ criterion. \cite{Krueger:etal:2015} proposed an improved $\CV$ procedure which uses nonparametric sequential test to speed up the conventional $\CV$. Moreover, \cite{Liu:etal:fastCV:2020} proposed a fast $\CV$ for kernel-based regression based on the Bouligand influence function ($\BIF$), which also requires the explicit expressions of $\BIF$. However, these existing methods cannot be applied in our context, where the data structures, loss functions, and candidate estimators are more complicated. Also, there is no immediate theory for nonparametric sequential tests under non-independent situations nor explicit expression for $\BIF$.

In this section, we assume that $\wh\f_{\s,i}$ is indexed by $\wh\bth_\s$, i.e., $\wh\f_{\s,i}=\f_{\s,i}(\wh\bth_\s)$, where $\f_{\s,i}(\bth_\s)$ is an $n_i$-dimensional vector valued function, $\wh\bth_\s$ is the a $q_s$-dimensional vector of estimators for $\bth_\s$, obtained by solving the estimating equation
$\U_\s(\bth_\s\mid\calD)=0_{q_s\times 1}$, and $q_s$ is allowed to diverge as $n\to\infty$ and satisfies $\max_{1\le s\le S}q_s/n\to0$. It is also worthwhile noting that although we use the term ``parameter'', our framework can also be used to nonparametric or semiparametric regression. The leave-one/subject-out counterpart of $\wh\f_{\s,i}$, say $\wh\f_{\s,i}^\ii$, can now be expressed as $\wh\f_{\s,i}^\ii=\f_{\s,i}(\wh\bth_{\s,\ii})$, where $\wh\bth_{\s,\ii}$ is the solution of $\U_\s(\bth_\s\mid\calD_\ii)=0_{q_s\times 1}$ and $\calD_\ii = \calD\setminus\calD_i$. We refer to $\wh\bth_{\s,\ii}$ as the conventional leave-one/subject-out estimator. The procedure is repeated for $n\times S$ times to obtain the conventional $\calC_n(\w)$, which is typically a time consuming task when the expression of $\U_\s(\cdot\mid\cdot)$ is complicated.

Now we develop a fast algorithm to overcome the computational bottleneck. To save notations, we suppress the use of $\calD$ or $\calD_\ii$ in the estimating equations and denote $\U_\s(\bth_\s\mid\calD)=\U_\s(\bth_\s)$ and $\U_\s(\bth_\s\mid\calD_\ii)=\U_{\s,\ii}(\bth_\s)$, respectively. Moreover, we denote $\u_{\s,i}(\bth_\s)=\U_\s(\bth_\s)-\U_{\s,\ii}(\bth_\s)$. In the remaining part of the current study, we assume that the up to third order partial derivatives of $\U(\cdot,\bth_\s)$ are continuous functions of $\bth_\s$ for each $s$. Let
\be
\J_\s(\bth_\s) = -\frac{\partial\U_\s(\bth_\s)}{\partial\bth_\s\tt},\quad\J_{\s,\ii}(\bth_\s) = -\frac{\partial\U_{\s,\ii}(\bth_\s)}{\partial\bth_\s\tt},\n
\ee
\be
\H_{\s,\ii}(\bth_\s) = \left\{\frac{\partial\J_{\s,\ii}(\bth_\s)}{\partial\theta_{\s,1}},\ldots,\frac{\partial\J_{\s,\ii}(\bth_\s)}{\partial\theta_{\s,q_s}}\right\}\n
\ee
and
\be
\V_{\s,\ii}(\bth_\s) = \left\{\frac{\partial\H_{\s,\ii}(\bth_\s)}{\partial\theta_{\s,1}},\ldots,\frac{\partial \H_{\s,\ii}(\bth_\s)}{\partial\theta_{\s,q_s}}\right\}.\n
\ee
Now, by the definition of $\wh\bth_{\s,\ii}$ and expanding $\U_{\s,\ii}(\wh\bth_{\s,\ii}) $ around $\wh\bth_\s$ up to the third-order, one has
\be
&\0_{q_s\times 1} &=\U_{\s,\ii}(\wh\bth_{\s,\ii})\n\\
&&=\U_\s(\wh\bth_\s)-\u_{\s,i}(\wh\bth_\s)+\J_{\s,\ii}(\wh\bth_\s)(\wh\bth_{\s}-\wh\bth_{\s,\ii})-\frac{\H_{\s,\ii}(\wh\bth_\s)}{2}(\wh\bth_{\s}-\wh\bth_{\s,\ii})^{\otimes2}\n\\
&&\quad+\frac{\V_{\s,\ii}(\bar\bth_\s)}{6}(\wh\bth_{\s}-\wh\bth_{\s,\ii})^{\otimes3}\n\\
&&=-\u_{\s,i}(\wh\bth_\s)+\J_{\s,\ii}(\wh\bth_\s)(\wh\bth_{\s}-\wh\bth_{\s,\ii})-\frac{\H_{\s,\ii}(\wh\bth_\s)}{2}(\wh\bth_{\s}-\wh\bth_{\s,\ii})^{\otimes2}\n\\
&&\quad+\frac{\V_{\s,\ii}(\bar\bth_\s)}{6}(\wh\bth_{\s}-\wh\bth_{\s,\ii})^{\otimes3},\n
\ee
where $\v^{\otimes k} = \underbrace{\v\otimes \v \otimes\ldots\otimes\v}_{k \text{ terms}}$ for any generic vector $\v$ and $\bar\bth_\s$ lies on the line segment joining $\wh\bth_{\s}$ and $\wh\bth_{\s,\ii}$. Then, after dropping the remainder term, we obtain the following estimating equation for solving the approximated leave-one/subject-out estimator:
\be
\bpsi_{\s,i}(\bth_s)=-\u_{\s,i}(\wh\bth_\s)+\J_{\s,\ii}(\wh\bth_\s)(\wh\bth_{\s}- \bth_\s)-\frac{\H_{\s,\ii}(\wh\bth_\s)}{2}(\wh\bth_{\s}- \bth_\s)^{\otimes2}.\label{eq:2nd:cv}
\ee
Let
$\wt\J_{\s,\ii}(\bth_\s)=\J_{\s,\ii}(\wh\bth_\s)-\H_{\s,\ii}(\wh\bth_\s)\left\{\I_{q_s}\otimes (\wh\bth_\s-\bth_\s)\right\}$.
It is readily seen that $\partial\bpsi_{\s,i}(\bth_\s)/\partial\bth_\s\tt = -\wt\J_{\s,\ii}(\bth_\s)$. The corresponding Newton-Raphson iterative algorithm for solving $\bpsi_{\s,i}(\bth_\s) = \0_{q_s\times1}$ is given by
$
\wt\bth_{\s,\ii}^{(d+1)} = \wt\bth_{\s,\ii}^{(d)}+\wt\J_{\s,\ii}^{-1}(\wt\bth_{\s,\ii}^{(d)})\bpsi_{\s,i}(\wt\bth_{\s,\ii}^{(d)}), \n
$
for $d=0,1,\ldots$ and we set the starting value as $\wt\bth_{\s,\ii}^{(0)}=\wh\bth_\s$. In practice, to avoid taking matrix inverse in each step of iteration, following  Equation (4.58) of \cite{Jiang:2007}, we adopt a second-order matrix approximation to further boost the algorithm, i.e.,
$
\wt\J_{\s,\ii}^{-1}(\wt\bth_{\s,\ii}^{(d)}) \approx  \J_\s^{-1}(\wh\bth_\s) + \J_\s^{-1}(\wh\bth_\s)\left\{\J_\s(\wh\bth_\s) -\wt\J_{\s,\ii}(\wt\bth_{\s,\ii}^{(d)})\right\} \J_\s^{-1}(\wh\bth_\s)
$.
In general, solving $\bpsi_{\s,i}(\bth_\s) = \0_{q_s\times1}$ is much easier than solving $\U_{\s,\ii}(\bth_\s) = \0_{q_s\times1}$. First, the former is a linear-quadratic equation system, and the latter is a general nonlinear equation system, which is typically complicated in the examples considered in the current study. Second, by adopting $\bpsi_{\s,i}(\cdot)$, we avoid involving $\calD_\ii$ directly. Instead, we need just to update the estimates based on the information embedded in $\wh\bth_\s$, $\J_{\s,\ii}(\wh\bth_\s)$, $\H_{\s,\ii}(\wh\bth_\s)$ and $\u_{\s,i}(\wh\bth_\s)$, which are readily available when the algorithm starts.

Let $\wt\bth_{\s,\ii}$ be the solution of $\bpsi_{\s,i}(\bth_s)=\0_{q_s\times1}$. In principle, $\bpsi_{\s,i}(\bth_s)$ approximates $\U_{\s,\ii}(\bth_\s)$ up to the second-order, and thus we term $\wt\bth_{\s,\ii}$ as the SEcond-order-Approximated Leave-one/subject-out ($\SEAL$) estimator and it is also expected to yield promising approximation to its conventional counterpart $\wh\bth_{\s,\ii}$. This conjecture is verified in Theorem S.1 of the
Supplementary Materials. In Theorem S.1, in terms of $l_{2k}$ norm ($l_k$ norm of a general random variable $z$ refers to $\E^{1/k}|z|^k$), Equation (S.2) indicates that the difference between $\wt\bth_{\s,\ii}$ and $\wh\bth_{\s,\ii}$ is at most of order $O(q_s^{3/2}/{n^2})$ and the result in Equation (S.3) implies that if a further smoothness condition is imposed, we can improve this difference to order $O(q_s^{5/2}/{n^3})$.

\subsection{$\MACV$ based on the $\SEAL$ estimator}
\label{subsec:MA:Est2nd}
The $\SEAL$ estimator $\wt\bth_{\s,\ii}$ leads to promising approximation to $\wh\bth_{\s,\ii}$ and is computationally efficient, which allows us to design a computationally attractive weight choice criterion based on it. Now consider a $\SEAL$ based weight choice criterion
$
\wt\calC_n(\w) =-2 \sumi Q\left\{\Y_i, \wt\f_{i,\ii}(\w)\right\}\n
$,
where $\wt\f_{i,\ii}(\w) =\sums w_s\f_{\s,i}(\wt\bth_{\s,\ii})$. Let  $
\wt\w = \argmin_{\w\in\calW}\wt\calC_n(\w)$. The convergence properties of $\wt\calC_n(\w)/n$ and $\wt\w$ are investigated in Corollary S.1 of the Supplementary Materials. This corollary indicates that the weight choice criterion evaluated at $\SEAL$ method has exactly the same limiting properties as that under conventional leave-subject/one-out method.  We now study the asymptotic optimality and consistency of $\wt\w$.

\begin{corollary}[\bf Asymptotic optimality under divergent $S$ with $\SEAL$ estimator]
	\label{coro:opt:MACV:2nd:div}
Assume that $S^{1/4}\{\bar q^{1/2}n^{1/2}/\xi_n + n/(b_n\xi_n)\}=o(1)$, $\{\log\log(n_\alpha)+\log(n/\xi_n)\}Sn^2/(n_\alpha\xi_n^2)=o(1)$, $\{\log\log(n_\alpha)+\log(n/\xi_n)\}S^{5/4}n/(n_\alpha^{3/4}\xi_n)=o(1)$, ${n}/{(S n_\alpha^2)}=o(1)$ and $\nu_n/\xi_n=o(1)$, as $n\to\infty$.
Under Assumptions S.2 and S.5--S.8 of the Supplementary Materials, for each $\w\in\calW$, assume that $z_i(\w)$'s satisfy the mixing condition given in (S.5), we have, as $n\to\infty$
	$
	{
		R_n(\wt \w)}/{\inf_{\w\in\calW}R_n(\w)}\to 1
	$,
	in probability.
\end{corollary}

The proof of Corollary \ref{coro:opt:MACV:2nd:div} is provided in Section S.2.9
of the
Supplemental Materials. In this corollary, we impose some requirements regarding the relationships between $\xi_n$, $n$, $S$ and $\bar q$, which can be viewed as a specific case of Assumption S.4 when $a_n=n^{1/2}/\bar q^{1/2}$.  Corollary \ref{coro:opt:MACV:2nd:div} indicates that the weight determined by $\SEAL$ based weight choice criterion is also asymptotically optimal.

\begin{corollary}[\bf Consistency of $\wt\w$ in identifying the exactly correct models]
	\label{coro:consist:MACV:2nd:div}
If Assumption \ref{con:dQ2} and Assumptions S.2, S.5--S.8 of the Supplementary Materials are satisfied, and for each $\w\in\calW$, $z_i(\w)$'s satisfy the mixing condition given in (S.5), then we have $\sum_{s\in D}\wt w_s \to 1$ in probability as
	$n\to\infty$, where $\wt w_s$ is the $s$th entry of $\wt\w$.
\end{corollary}

Along the same line as the proof of Theorem \ref{th:consist:MACV}, we can prove Corollary \ref{coro:consist:MACV:2nd:div} by using the uniform convergence properties established in Corollary S.1 and is omitted. Corollary \ref{coro:consist:MACV:2nd:div} implies that our proposed $\SEAL$ based criterion is asymptotically equivalent to the conventional $\MACV$ in identifying the correct models when at least one of the candidate models is correctly specified in the exact sense.

\section{Simulation studies} \label{sec:sim:miss}

In this section, we conduct simulation experiments to assess the finite sample prediction accuracy of the proposed $\SEAL$-based $\MACV$ (denoted as $\MA_\SEAL$) and compare it with some competing methods. $\MACV$ based on conventional leave-subject/one-out estimator is not considered due to heavy computation cost. However, we compare the conventional method and $\SEAL$ in the case studies where the sample sizes are relatively small. To provide further information, we also consider the following quantity to measure the similarity between different candidate models,
$
 \hat\gamma  = {2}\sum\nolimits_{1 \le s < s_1 \le S} {\hat c_{s,s_1}}/\{{S( S-1 ){( {\sums \wh\sigma_s }/S)}^2}\}
$,
where $\hat c_{s,s_1}$ is the sample covariance between the losses by the $s$th and $s_1$th candidate models and $\wh\sigma_s$ is the sample standard deviation of the losses by the $s$th model among $M$ replications. Similar measure was proposed in \cite{Breiman:2001} to quantify the similarity between different trees in random forest. We also consider the overall performance measure $\bar L_M$, which is the average of the losses from all the $S$ candidate models among $M$ replications.
Further simulation studies under different data settings ($\GEE$ models with high dimensional covariates, within-cluster correlation structure modeling, conditional prediction in spatial data, and quantile regression with a potential correlation structure) are reported in Sections S.4.1. The power of $\MACV$ in identifying the exactly correct models is also assessed via simulation in Section S.4.2.
\renewcommand{\theDesign}{\arabic{Design}}

\setcounter{Design}{0} 

\begin{Design}[Longitudinal data with discrete responses]
	\label{sim:miss:GEE} The purpose of the current simulation study is to assess the performance of the proposed $\MA_\SEAL$ in predicting discrete longitudinal data. We employ $\GEE$ to conduct parameter estimation via MATLAB toolbox GEEQBOX \citep{Ratcliffe:Shults:2008} and compare $\MA_\SEAL$ with its competitors when all the candidate $\GEE$s are misspecified (specially, the marginal mean function). We examine the following competing methods: model averaging with equal weights (denoted as $\Equal$), model selection methods based on $\CV$ (denoted by $\CV$, which is also based on $\SEAL$ estimator), the quasi-likelihood under the independence model criterion ($\QIC$) of \cite{Pan:2001} (denoted by $\QIC_\Pan$) and modified $\QIC$ by \cite{Imori:2015} (denoted by $\QIC_\Imori$). The data are generated by the algorithm based on the conditional linear family \citep{Qaqish:2003}. The binary responses $Y_{ij}$'s ($i=1,\ldots,n;j=1,\ldots,4$) have the marginal mean of form $p_{ij} = 1/\left\{1+\exp(-\x_{ij}\tt\bb)\right\}$, $\x_{ij} = (1,x_{ij,1},\ldots,x_{ij,p})\tt$, $x_{ij,k}$'s ($k=1,\ldots,p$) are independent and identically distributed as $\Normal(0,1)$.
$\bb$ takes value from $\{(0.2, 0, 0, 0, -0.5,  0.1)\tt, (0.2, 0, 0, 0, -0.5,  0.3, -0.1),(0.2, 0, 0, 0, 0,-0.5,  0.3, -0.1)\tt\}$ for $n=\{100,150,200\}$, respectively.
 Moreover, we set $\corr(Y_{ij},Y_{il}) = \rho^{|j-l|}$ with $\rho\in\{0, 0.3\}$, where we set the largest value of $\rho$ as $0.3$ to guarantee the natural restriction condition \citep{Qaqish:2003}. Here, $\Y_i$'s are set to be independent for different $i$ and the true within subject correlation structure is ``independent structure'' or ``first-order autoregressive correlation matrix'', depending on the value of $\rho$. In the candidate $\GEE$s, the logit-link function is adopted. For the configuration of linear predictors in candidate $\GEE$s, the constant term is always included, and $x_{ij,k}$'s
($k=1,\ldots,p-1$) are optional covariates that may or may not be included in the linear predictor. The last covariate $x_{ij,p}$ is deliberately dropped from all the candidate models so that the mean functions in all the candidate $\GEE$s are misspecified. Two types of working correlation structures are involved, i.e., ``equicorrelated structure'' and ``first-order autoregressive working correlation matrix'' and this configuration allows us to mimic the situation where the working correlation structures are subjected to possible misspecification. Therefore, we have $2\times 2^{p-1}$ candidate models and all of them are misspecified.
Under the current setup, there are at least two models contain the same number of unknown parameters, therefore the notion of the full-model becomes ambiguous. This is another reason that motivates us to consider model averaging in the current scenario.
\begin{table}[pbth]
	\small
\renewcommand{\arraystretch}{0.58}
\tabcolsep=.08cm
\caption{Results for Design \ref{sim:miss:GEE}: sample-based mean,
		$25\%$, $50\%$ and $75\%$ quantiles of scaled loss (scaled by $10\times n^{-1}$) by different methods}{
   \begin{tabular}{ccccccccrccccc}
\hline
          &              \multicolumn{12}{c}{Correlated binary data}         &        \\
          &       & \multicolumn{6}{c}{$\rho=0.3$}                   &       & \multicolumn{5}{c}{$\rho=0$} \\
    \multicolumn{1}{l}{$n$} & \multicolumn{1}{l}{$S$} &       & \multicolumn{1}{l}{$\MA_\SEAL$} & \multicolumn{1}{l}{$\Equal$} & \multicolumn{1}{l}{$\CV$} & \multicolumn{1}{l}{$\QIC_\Imori$} & \multicolumn{1}{l}{$\QIC_\Pan$} &       & \multicolumn{1}{l}{$\MA_\SEAL$} & \multicolumn{1}{l}{$\Equal$} & \multicolumn{1}{l}{$\CV$} & \multicolumn{1}{l}{$\QIC_\Imori$} & \multicolumn{1}{l}{$\QIC_\Pan$} \\
   \cline{4-8}\cline{10-14}
   \multirow{6}[0]{*}{100} & \multirow{6}[0]{*}{32} & mean  & 0.112  & \multicolumn{1}{r}{0.221 } & 0.122  & 0.122  & 0.123  &       & 0.099  & 0.211  & 0.114  & \multicolumn{1}{r}{0.115 } & \multicolumn{1}{r}{0.115 } \\
          &       & 25\%  & 0.057  & \multicolumn{1}{r}{0.173 } & 0.056  & 0.055  & 0.058  &       & 0.050  & 0.163  & 0.047  & \multicolumn{1}{r}{0.048 } & \multicolumn{1}{r}{0.049 } \\
          &       & 50\%  & 0.094  & \multicolumn{1}{r}{0.210 } & 0.105  & 0.101  & 0.104  &       & 0.080  & 0.204  & 0.094  & \multicolumn{1}{r}{0.096 } & \multicolumn{1}{r}{0.095 } \\
          &       & 75\%  & 0.149  & \multicolumn{1}{r}{0.259 } & 0.167  & 0.166  & 0.168  &       & 0.126  & 0.247  & 0.154  & \multicolumn{1}{r}{0.154 } & \multicolumn{1}{r}{0.155 } \\
         \cline{4-8}\cline{10-14}
          &       & $\wh\gamma$ &       &       & 0.771  &       &       &       &       &       & 0.669  &       &  \\
          &       & $\bar L_M$  &       &       & 0.383  &       &       &       &       &       & 0.373  &       &  \\
\cline{3-14}
    \multirow{6}[0]{*}{150} & \multirow{6}[0]{*}{64} & mean  & 0.100  & \multicolumn{1}{r}{0.263 } & 0.103  & 0.104  & 0.104  &       & 0.094  & 0.256  & 0.101  & \multicolumn{1}{r}{0.102 } & \multicolumn{1}{r}{0.102 } \\
          &       & 25\%  & 0.057  & \multicolumn{1}{r}{0.222 } & 0.057  & 0.055  & 0.058  &       & 0.054  & 0.215  & 0.052  & \multicolumn{1}{r}{0.053 } & \multicolumn{1}{r}{0.053 } \\
          &       & 50\%  & 0.086  & \multicolumn{1}{r}{0.256 } & 0.089  & 0.088  & 0.090  &       & 0.080  & 0.251  & 0.085  & \multicolumn{1}{r}{0.086 } & \multicolumn{1}{r}{0.088 } \\
          &       & 75\%  & 0.129  & \multicolumn{1}{r}{0.301 } & 0.134  & 0.134  & 0.134  &       & 0.118  & 0.293  & 0.134  & \multicolumn{1}{r}{0.135 } & \multicolumn{1}{r}{0.135 } \\
          \cline{4-8}\cline{10-14}
          &       &$\wh\gamma$  &       &       & 0.725  &       &       &       &       &       & 0.653  &       &  \\
          &       &$\bar L_M$   &       &       & 0.466  &       &       &       &       &       & 0.462  &       &  \\
  \cline{3-14}
    \multirow{6}[0]{*}{200} & \multirow{6}[0]{*}{128} & mean  & 0.088  & \multicolumn{1}{r}{0.260 } & 0.094  & 0.093  & 0.094  &       & 0.081  & 0.252  & 0.090  & \multicolumn{1}{r}{0.091 } & \multicolumn{1}{r}{0.091 } \\
          &       & 25\%  & 0.053  & \multicolumn{1}{r}{0.223 } & 0.054  & 0.054  & 0.055  &       & 0.049  & 0.219  & 0.048  & \multicolumn{1}{r}{0.049 } & \multicolumn{1}{r}{0.049 } \\
          &       & 50\%  & 0.078  & \multicolumn{1}{r}{0.257 } & 0.084  & 0.082  & 0.085  &       & 0.069  & 0.249  & 0.079  & \multicolumn{1}{r}{0.079 } & \multicolumn{1}{r}{0.079 } \\
          &       & 75\%  & 0.111  & \multicolumn{1}{r}{0.292 } & 0.122  & 0.121  & 0.123  &       & 0.100  & 0.282  & 0.118  & \multicolumn{1}{r}{0.119 } & \multicolumn{1}{r}{0.119 } \\
          \cline{4-8}\cline{10-14}
          &       & $\wh\gamma$  &       &       & 0.718  &       &       &       &       &       & 0.644  &       &  \\
          &       & $\bar L_M$   &       &       & 0.459  &       &       &       &       &       & 0.455  &       &  \\
 \hline
          &            \multicolumn{12}{c}{Correlated count data}              &  \\
  \multirow{6}[0]{*}{100} & \multirow{6}[0]{*}{32}    & mean  & 0.222  & 0.980  & 0.234  & 0.234  & 0.235  &       & 0.207  & 0.966  & 0.223  & 0.225  & 0.225  \\
          &       & 25\%  & 0.165  & 0.881  & 0.169  & 0.166  & 0.170  &       & 0.160  & 0.870  & 0.160  & 0.163  & 0.162  \\
          &       & 50\%  & 0.201  & 0.974  & 0.213  & 0.212  & 0.214  &       & 0.188  & 0.961  & 0.204  & 0.206  & 0.208  \\
          &       & 75\%  & 0.257  & 1.070  & 0.278  & 0.279  & 0.278  &       & 0.235  & 1.057  & 0.264  & 0.267  & 0.267  \\
          \cline{4-8}\cline{10-14}
          &       & $\wh\gamma$ &       &       & 0.775  &       &       &       &       &       & 0.650  &       &  \\
          &       & $\bar L_M$    &       &       & 1.818  &       &       &       &       &       & 1.804  &       &  \\
             \cline{3-14}
   \multirow{6}[0]{*}{150} & \multirow{6}[0]{*}{64}    & mean  & 0.220  & 1.447  & 0.225  & 0.225  & 0.225  &       & 0.210  & 1.434  & 0.218  & 0.219  & 0.219  \\
          &       & 25\%  & 0.177  & 1.340  & 0.177  & 0.177  & 0.178  &       & 0.171  & 1.334  & 0.173  & 0.174  & 0.173  \\
          &       & 50\%  & 0.206  & 1.442  & 0.214  & 0.213  & 0.214  &       & 0.198  & 1.432  & 0.206  & 0.207  & 0.207  \\
          &       & 75\%  & 0.253  & 1.542  & 0.257  & 0.259  & 0.258  &       & 0.235  & 1.530  & 0.249  & 0.251  & 0.252  \\
           \cline{4-8}\cline{10-14}
          &       & $\wh\gamma$ &       &       & 0.746  &       &       &       &       &       & 0.663  &       &  \\
          &       & $\bar L_M$    &       &       & 2.774  &       &       &       &       &       & 2.765  &       &  \\
              \cline{3-14}
  \multirow{6}[0]{*}{200} & \multirow{6}[0]{*}{128}   & mean  & 0.213  & 1.433  & 0.219  & 0.220  & 0.220  &       & 0.203  & 1.405  & 0.213  & 0.214  & 0.214  \\
          &       & 25\%  & 0.180  & 1.348  & 0.181  & 0.181  & 0.181  &       & 0.175  & 1.327  & 0.178  & 0.178  & 0.178  \\
          &       & 50\%  & 0.202  & 1.429  & 0.207  & 0.208  & 0.207  &       & 0.194  & 1.404  & 0.204  & 0.205  & 0.205  \\
          &       & 75\%  & 0.235  & 1.519  & 0.245  & 0.246  & 0.245  &       & 0.221  & 1.479  & 0.239  & 0.239  & 0.240  \\
           \cline{4-8}\cline{10-14}
          &       & $\wh\gamma$ &       &       & 0.721  &       &       &       &       &       & 0.636  &       &  \\
          &       & $\bar L_M$    &       &       & 2.642  &       &       &       &       &       & 2.635  &       &  \\

          \hline
    \end{tabular}%
}
	\label{tab:sim:miss:GEE:bi}
\end{table}
The loss function defined in Equation (\ref{LBernoulli}) is employed to assess
the prediction accuracy of different methods. Correlated counts $Y_{ij}$'s are also generated, where $\E(Y_{ij})=\exp(\x_{ij}\tt\bb)$ and log-link function is adopted. The loss function in Equation (\ref{LPoisson}) is considered.  Other configurations are same as those in the binary data.

In each parameter setting, $M=1000$ replications are generated.
 Sample-based mean and $25\%$, $50\%$ and $75\%$ quantiles of scaled losses
(scaled by $10\times n^{-1}$) for different methods under various
combinations of sample sizes and $\rho$ are reported in Table \ref{tab:sim:miss:GEE:bi}, together with $S$, $\wh\gamma$ and $\bar L_M$.
It is observed that in terms of sample-based mean, $50\%$ and $75\%$ quantiles of the losses,  the proposed model averaging method consistently outperforms its competitors in all the settings.
As for $25\%$ quantiles of the losses, all methods perform closely.
Moreover, $\CV$, $\QIC_{\Imori}$, and $\QIC_{\Pan}$ lead to very similar performances
when the sample size becomes larger. Equal weight model averaging leads to the worst performance, which is probably due to the reason that the strategy puts equal weight on all the models and thus tends to be impacted by models with very large losses. In fact, under the scenario of binary responses, all the $\bar  L_M$'s are larger than $0.37$, which is at least four times larger than the average loss of the best method. Whereas for the scenario of count data, $\bar  L_M$'s are at least eight times larger than the average loss of the best method. These observations indicate the presence of very poor models. We also find that in comparison to model selection, the advantage of $\MA_\SEAL$ is more pronounced when $\wh\gamma$ remains at a lower level (e.g., when $\rho=0$). The reason is that if all the candidate models have similar performances, it is relatively easier for the model selection criterion to identify the best models.
\end{Design}

\section{Case study}
\label{sec:case:resp}
In this section, we apply $\MACV$ (both conventional and $\MA_\SEAL$) to analyze a dataset from a respiratory study \citep{Davis:1991}.
The trial of respiratory study involved $n=111$ participants (subjects) and each subject has five observations (month 0, month 1, $\ldots$, month 4).
Following \cite{Everitt:Hothorn:2010}, we take the status for month 0 as baseline status and
use it as a covariate. Thus the rearranged data have 4 observations for each subject.

The response variable is the status of subject (good $=1$, poor $=0$). Due to the
response variable's binary nature, GEE with the logit-link
is applied to fit the data. Moreover, a constant term is always included in the linear predictor of all the candidate GEEs.
Other optional covariates are Center (center 1 $=0$, center 2 $=1$),
Treat (treatment $=1$, placebo $=0$), Sex (male $=1$, female $=0$), Baseline (good $=1$, poor $=0$) and Age.
These optional covariates may or may not be included in the candidate GEEs.  We also consider two working correlation
structures, i.e., ``equicorrelated structure'' and ``first-order autoregressive working correlation matrix''. Therefore, we have $S=2\times 2^5 = 64$ candidate models.
To assess the performance of the proposed $\MACV$ method
and its competitors ($\CV$, $\QIC_{\Imori}$ and $\QIC_{\Pan}$) {in predicting the new subjects}, we randomly group the full dataset into training dataset
and test set and repeat this process for $M=1000$ times.
The proportion of test set $r_{\test}$ takes value $0.7$ or $0.3$.
$n_{\test} = floor(n\times r_{\test})$ is the number of subjects in the test set.
Parameter estimation and model averaging (or selection) are
based on training data. Stimulated by the loss function given in (\ref{LBernoulli}),
in the $m$th replication ($m=1,\ldots,1000$), we employ the following
scaled empirical loss to measure the prediction accuracy of a method.
\bse
\wh L_{(m)}
=-\frac{2}{n_{\test}} \sum\limits_{i=1}^{n_{\test}}\sum\limits_{j=1}^{4} \left\{Y_{ij,\test}^{(m)}
\log\left(\displaystyle\frac{\wh\mu_{ij,\train}^{(m)}}{1-\wh\mu_{ij,\train}^{(m)}}
\right)
+ \log(1 - \wh\mu_{ij,\train}^{(m)})\right\}-\wh L_\ast^{(m)},\n
\ese
where $Y_{ij,\test}^{(m)}$ is the binary status of the $j$th observation of the $i$th subject in the
test data,
$\wh\mu_{ij,\train}^{(m)}$ is the prediction of
$\E(Y_{ij,\test}^{(m)})$ by different methods ($\MACV$, CV, $\QIC_{\Imori}$ or $\QIC_{\Pan}$) based on the training data, and
{\small
\bse
\wh L_\ast^{(m)}
=\min_{1\le s\le S} \left[-\frac2
{n_{\test}} \sum_{i=1}^{n_{\test}}\sum_{j=1}^{4} \left\{Y_{ij,\test}^{(m)}
\log\left(\displaystyle\frac{\wh\mu_{\s,ij,\train}^{(m)}}{1-\wh\mu_{\s,ij,\train}^{(m)}}\right)
+ \log(1 - \wh\mu_{\s,ij,\train}^{(m)})\right\}\right],
\ese
}
with $\wh\mu_{\s,ij,\train}^{(m)}$ being the  prediction of
$\E(y_{ij,\test}^{(m)})$ by the $s$th candidate GEE.
The empirical loss
$\wh L_{(m)}$ finds its root in Akaike information or relative Kullback-Leibler divergence, and we can use
it to measure the divergence between the future data (test data) and predictions based on training data.
A similar measure has also been advocated by \cite{Zhang:Yu:Zou:Liang:2016} and \cite{Yu:Zhang:Yau:2018}.

Sample-based mean, $25\%$, $50\%$ and $75\%$ quantiles of $\wh L_{(m)}$'s ($m=1,\ldots,1000$)
are reported in Table \ref{tab:case:resp}, where we also report the computing time (seconds) of $\MA_\SEAL$ and  conventional $\MACV$ (denoted by $\MA_\CONVEN$). It is observed that in terms of prediction loss, $\MA_\SEAL$ and conventional $\MACV$ yield almost identical performance, while $\MA_\SEAL$ costs much shorter computing time. It can also be observed from Table \ref{tab:case:resp} that
when the sample size of training data is smaller ($r_{\test} = 0.7$), $\MA_\SEAL$ and conventional $\MACV$ methods substantially outperform their competing methods ($\Equal$, $\CV$, $\QIC_{\Imori}$ and $\QIC_{\Pan}$). Moreover, when the sample size in
the training data becomes larger ($r_{\test} = 0.3$), the situation becomes slightly mixed. The $\wh L_{(m)}$'s  ($m=1,\ldots,1000$) by $\MACV$ method have the smallest sample mean, median and $75\%$ quantile, whereas those by $\QIC_{\Imori}$ have the smallest $25\%$ quantile.
One possible explanation of this phenomenon is that when the sample size is smaller, the model selection
procedure induces a higher level of sampling error to the post-model-selection-prediction, and such an error can be reduced by $\MACV$ method, by integrating the predictions from different candidate models. When the sample size becomes larger, such sampling error caused by model selection becomes smaller and poses a less severe impact on prediction. These findings further support the use of our $\MACV$ method in practical situations. It can also be observed that the when the training sample size increases from $33$ (i.e., $r_{\test}=0.7$) to $77$ (i.e., $r_{\test}=0.3$), the computing time of $\SEAL$ estimator only increases by $22\%$, whereas that of conventional leave-subject-out estimator increases for about $155\%$, demonstrating the computational efficiency of $\SEAL$ in practice.

 We also analyze the epilepsy seizure count data, neighborhood crimes data and airline data in Section Section S.4.3 of the
 Supplemental Materials. These case studies further verify the advantage of the proposed $\MACV$ over its competitors and demonstrate the computational efficiency of $\SEAL$ in practical situations.

\begin{table}[pbht]
\small
\renewcommand{\arraystretch}{0.58}
\tabcolsep=.05cm
\caption{Results for the respiratory study: sample based mean
	and  $25\%$, $50\%$ and $75\%$ quantiles of $\wh L_{(m)}$'s ($ m=1,\ldots,1000$) and computing time (seconds) by different methods,  $r_{\test}\in\{0.7,0.3\}$\label{tab:case:resp}}
\centering{
    \begin{tabular}{cccccccccccccc}
    \hline
          & \multicolumn{6}{c}{$r_{\test}=0.7$}                       &       & \multicolumn{6}{c}{$r_{\test}=0.3$} \\
          \cline{2-7}\cline{9-14}
              & $\MA_\SEAL$ & $\MA_\CONVEN$ & $\Equal$ & $\CV$    & $\QIC_\Imori$ & $\QIC_\Pan$   &       & $\MA_\SEAL$ & $\MA_\CONVEN$ & $\Equal$ & $\CV$    & $\QIC_\Imori$ & $\QIC_\Pan$  \\

    mean  & 0.237  & 0.238  & 0.249  & 0.479  & 0.502  & 0.511  &       & 0.196  & 0.196  & 0.340  & 0.233  & 0.228  & 0.239  \\
    25\%  & 0.066  & 0.066  & 0.122  & 0.069  & 0.097  & 0.091  &       & 0.094  & 0.094  & 0.179  & 0.105  & 0.067  & 0.098  \\
    50\%  & 0.175  & 0.175  & 0.259  & 0.210  & 0.255  & 0.268  &       & 0.164  & 0.164  & 0.355  & 0.168  & 0.165  & 0.173  \\
    75\%  & 0.317  & 0.317  & 0.391  & 0.486  & 0.529  & 0.566  &       & 0.252  & 0.253  & 0.496  & 0.286  & 0.317  & 0.303  \\
       Time  & 1837.14    & 	5996.49    &     --  &   --    &    --   &   --    &       &    2241.29 	   & 15284.27    &  --     &    --   &    --   & -- \\

   \hline
    \end{tabular}%
}
\end{table}

\section{Discussions}\label{sec:conclude}

In this paper, we have proposed a model averaging method in a unified framework based on $\CV$. The new unified framework is flexible for different types of data and broad loss functions. Within this framework, we exemplify four new optimal model averaging estimators under four important situations, i.e., longitudinal data with discrete responses, within-cluster correlation structure modeling,  conditional prediction in spatial data, and quantile regression with a potential correlation structure. We also propose $\SEAL$ to reduce the computational burden.
The unified framework covers many existing model averaging estimators such as jackknife model averaging for linear models \citep{hansen.racine:2012,zhang.wan.ea:2013}, longitudinal data models \citep{Gao:Zhang:Wang:Zou:2016}, generalized additive partial linear models \citep{Chen:etal:2023} and quantile regression \citep{lu.su:2014,Wang:etal:2023}. The proposed $\MACV$ enjoys excellent theoretical properties. Simulation and cases studies showed that the $\MACV$ has promising finite sample performance under different situations. The technique used in the theoretical development and analysis is flexible to different data structures, loss functions, and estimation methods. Thus we expect that our method can be potentially applied in the areas of complex data analysis such as high-dimensional smoothed quantile regression \citep{Tan:Wang:Zhou:2022} and generalized additive model with penalized spline \citep{Hui:etal:2019} and so on.

\section*{Acknowledgements}
The authors would like to thank the Co-Editor, Professor Annie Qu, an associate editor, and two anonymous reviewers for their constructive suggestions and comments, which substantially improved the earlier version of this article.

{
\spacingset{1.75}

}

\end{document}